\documentclass{article}
\usepackage{dcase2019_techrep,amsmath,graphicx,url,times,booktabs, tabularx}
\usepackage{float}

\title{Acoustic Scene Classification Using Fusion of Attentive Convolutional Neural Networks for DCASE2019 Challenge}

\name{Hossein Zeinali, Luk\'{a}\v{s} Burget and Jan ``Honza'' \v{C}ernock\'{y}}
\address{Brno University of Technology, Speech@FIT and IT4I Center of Excellence, Czech Republic}

\begin{document}

\ninept
\maketitle

\begin{sloppy}

\begin{abstract}
In this report, the Brno University of Technology (BUT) team submissions for Task 1 (Acoustic Scene Classification, ASC) of the DCASE-2019 challenge are described. Also, the analysis of different methods is provided. The proposed approach is a fusion of three different Convolutional Neural Network (CNN) topologies. The first one is a VGG like two-dimensional CNNs. The second one is again a two-dimensional CNN network which uses Max-Feature-Map activation and called Light-CNN (LCNN). The third network is a one-dimensional CNN which mainly used for speaker verification and called x-vector topology. All proposed networks use self-attention mechanism for statistic pooling. As a feature, we use a 256-dimensional log Mel-spectrogram. Our submissions are a fusion of several networks trained on 4-folds generated evaluation setup using different fusion strategies.
\end{abstract}


\section{Introduction}
\label{sec:intro}

This report describes Brno University of Technology (BUT) team submissions for the ASC challenge of DCASE 2019. We proposed three different deep neural network topologies for this task. The first one is a VGG like~\cite{simonyan2014very} two-dimensional CNN network for processing audio segments. The second network is again a 2-dimensional CNN network which called Light-CNN~\cite{wu2018light}. This network uses several Max-Feature-Map activations for reducing the number of channels after convolutional layers. Light-CNN was successfully used for spoofing attack detection challenge~\cite{Lavrentyeva2017}. We also used a fusion of this network with a VGG network for the last spoofing challenge~\cite{Zeinali2019spoofing}. The last network topology uses a one-dimensional CNN along the time axis. This topology is mainly used to extract fixed-length embeddings of (possibly variable length) acoustic segments. This architecture has been previously found useful for other speech processing tasks such as speaker recognition~\cite{snyder2018x, zeinali2019improve}, where the extracted embeddings were called x-vectors. In the previous DCASE challenge (i.e. DCASE 2018) we have used this network for both classification (i.e. like other two networks) and extracting x-vector embeddings~\cite{zeinali2018convolutional} while in this challenge we only use it for classification. All proposed networks were trained with 256-dimensional log Mel-spectrogram features. In all networks we use self-attention mechanism~\cite{zhu2018self,okabe2018attentive,chowdhury2017attention} for pooling instead of simple average pooling. Our submissions are based on fusions of different networks trained on the task development data.

The current ASC challenge has three sub-tasks: In task1a, participants are allowed to use only the fixed development data for training. Task1b is similar to task1a except that the test files are from different mobile channels. Finally, task1c is an open set classification challenge where the test recording may be from a different environment than the 10 predefined target classes, in which case it should be classified as "unknown". We have participated in task1a only.

\section{Dataset}

In this challenge, an enhanced version of ASC dataset was used~\cite{mesaros2018multi}. The dataset consists of recordings from 10 scene classes and was collected in 12 large European cities and in different environments in each city. The development set of the dataset for task1a consists of 1440 segments for each acoustic scene and in total 40 hours of audio. This part only contains the recordings from 10 cities. The evaluation set consists of 7200 audio segments and was collected in different location of these 10 cities as well as in two not-seen cities, to test the generalization properties of the systems. Each segment has an exactly 10-second duration, this is achieved by splitting longer audio recordings from each environment (between 5-6 minutes). The dataset includes a predefined validation fold. Each team can also create its own folds, so we have created a 4-folds cross-validation setup for system development. The audio segments are 2-channels stereo files, recorded at 48\:KHz sampling rate.

\section{Data Processing}

\subsection{Features}

The log Mel-scale spectrogram was used as a feature in this challenge. For extracting the features, first, we converted the audio to a mono-channel and removed the amplitude bias by subtracting the audio segment's mean from the signal. Then short time Fourier transform is computed on 2048 samples Hamming windowed frames with 430 samples overlap from downsampled signals to 22050\,Hz. Next, the power spectrum is transformed into 256 Mel-scale band energies and, finally, the log of these energies is taken. The features are extracted using librosa toolbox~\cite{mcfee2015librosa}.

\subsection{Example generation for network training}

The procedure for generating training examples can greatly affect the performance of neural networks in audio processing. Therefore, we experimented with several different strategies to find the best example generation method.  We randomly select a subpart of each audio segment as an example in the training time. Because we have an attention pooling layer for the time axis in the all proposed networks, we can have input with different size during the training and test time. Initially, we used four-second segments but after doing several experiments, we found that networks trained on the smaller segments performed better than those trained on large segments, mainly because they overfit less to the training data. The size of the examples used to train the submitted systems is only 128 frames from 512 extracted frames for whole 10 second segments.



\section{CNN topologies}

We have used three different CNN topologies for this challenge. The first one is a VGG like two-dimensional CNN. The second topology is an enhanced version of Light-CNN (LCNN) which used Max-Feature-Map (MFM) as an additional non-linearity. MFM reduces the number of kernels to half. As a result, the final network has fewer parameters and this is the main reason that this network called Light-CNN. The last network is a one-dimensional CNN topology known as x-vector which is the state-of-the-art method for speaker recognition~\cite{snyder2018x}. In all networks, we have used a self-attention mechanism in pooling layer instead of common average pooling. Self-attention can be considered as a weighted average (and weighted standard deviation). All networks are described in more detail in the following sub-sections.

\subsection{VGG-like network}

The VGG network comprises several convolutional and pooling layers followed by a statistics pooling and several dense layers which perform classification. Table~\ref{tab:vgg} provides a detailed description of the proposed VGG architecture. There are 6 convolutional blocks in the model, each containing 2 convolutional layers and one max-pooling. Each max-pooling layer reduces the size of the frequency axis to half while only one of them reduces the temporal resolution. After the convolutional layers, there is an attention pooling layer. This layer operates only on the time axis and calculates the weighted mean over time. This layer will be explained in more details in the following section. After this layer, there is a flatten layer which simply concatenates the 4 remaining frequency dimensions. Finally, there are 3 dense layers which perform the classification task.

\begin{table}[!t]
  \renewcommand{\arraystretch}{1.1}
	\centering
	\caption{\label{tab:vgg}The proposed VGG architecture. Conv2D: two  dimensional convolutional layer. AttentionPooling: a layer which calculate the weighted mean in time axis using attention mechanism and reduce the shape (remove the time axis). Dense: fully connected dense layer. N in the third column indicates the segment length which is 128 for the training phase and 512 for the evaluation phase. The attention layer here only uses the mean statistics.}
	\vspace{2mm}
	\setlength\tabcolsep{4pt}
	\begin{tabular}{l c c r}
		\toprule
		\toprule
		\textbf{Layer name}   & \textbf{Filter}   & \textbf{Output}                   & \textbf{\#Params} \\
		\midrule
		Input                 & --                & 256 $\times$ N $\times$ 1      & -- \\
		Conv2D-1-1            & 3 $\times$ 3      & 256 $\times$ N $\times$ 32     & 608 \\
		Conv2D-1-2            & 3 $\times$ 3      & 256 $\times$ N $\times$ 32     & 9.2K \\
		MaxPooling-1          & 2 $\times$ 1      & 128 $\times$ N $\times$ 32     & -- \\
		\midrule
		Conv2D-2-1            & 3 $\times$ 3      & 128 $\times$ N $\times$ 64     & 18.5K \\
		Conv2D-2-2            & 3 $\times$ 3      & 128 $\times$ N $\times$ 64     & 37K \\
		MaxPooling-2          & 2 $\times$ 1      & 64  $\times$ N $\times$ 64     & --  \\
		\midrule
		Conv2D-3-1            & 3 $\times$ 3      & 64 $\times$ N $\times$ 128     & 74K  \\
		Conv2D-3-2            & 3 $\times$ 3      & 64 $\times$ N $\times$ 128     & 148K \\
		MaxPooling-3          & 2 $\times$ 1      & 32 $\times$ N $\times$ 128     & --   \\
		\midrule
		Conv2D-4-1            & 3 $\times$ 3      & 32 $\times$ N $\times$ 256     & 295K  \\
		Conv2D-4-2            & 3 $\times$ 3      & 32 $\times$ N $\times$ 256     & 590K \\
		MaxPooling-4          & 2 $\times$ 1      & 16 $\times$ N $\times$ 256     & --   \\
		\midrule
		Conv2D-5-1            & 3 $\times$ 3      & 16 $\times$ N $\times$ 256     & 590K  \\
		Conv2D-5-2            & 3 $\times$ 3      & 16 $\times$ N $\times$ 256     & 590K \\
		MaxPooling-5          & 2 $\times$ 1      & 8  $\times$ N $\times$ 256     & --   \\
		\midrule
		Conv2D-6-1            & 3 $\times$ 3      & 8 $\times$ N $\times$ 256     & 590K  \\
		Conv2D-6-2            & 3 $\times$ 3      & 8 $\times$ N $\times$ 256     & 590K \\
		MaxPooling-6          & 2 $\times$ 1      & 4 $\times$ N $\times$ 256     & --   \\
		\midrule
		AttentionPooling      & --                & 4 $\times$ 256                 & 66K  \\
		Flatten               & --                & 1024                           & --  \\
		\midrule
		Dense1                & --                & 256                            & 262K  \\
		Dense2                & --                & 256                            & 66K  \\
		Dense (softmax)       & --                & 10                             & 2570  \\
		\midrule
		Total                 & --                & --                             & 3950K  \\
		\bottomrule
		\bottomrule
	\end{tabular}
	\vspace{-2mm}
\end{table}

\subsection{Light CNN (LCNN)}

Table~\ref{tab:lcnn} shows the used LCNN topology for this challenge. This network is a combination of convolutional and max-pooling layers and uses Max-Feature-Map (MFM) as an additional non-linearity. MFM is a layer which simply reduce the number of output channels to the half by taking the maximum of two consecutive channels (or any two channels, e.g. $i, \frac{N}{2} + i$). The rest of this network (statistics and classification parts) is identical to the proposed VGG network.

\begin{table}[!t]
	\centering
	\caption{\label{tab:lcnn} The proposed LCNN architecture. MFM: Max-Feature-Map activation. N in the third column indicates the segment length which is 128 for the training phase and 512 for the evaluation phase. The attention layer here only uses the mean statistics.}
	\vspace{2mm}
	\setlength\tabcolsep{4pt}
	\begin{tabular}{l c l r}
		\toprule
		\toprule
		\textbf{Layer name}   & \textbf{Filter}   & \textbf{Output}                   & \textbf{\#Params} \\
		\midrule
		Input                 & --                & 256 $\times$ N $\times$ 1       & -- \\
		Conv2D-1-1            & 5 $\times$ 5      & 256 $\times$ N $\times$ 32      & 832 \\
		MFM-1-1               & --                & 256 $\times$ N $\times$ 16      & -- \\
		MaxPooling-1          & 2 $\times$ 1      & 128 $\times$ N $\times$ 16      & -- \\
		\midrule
		Conv2D-2-1            & 1 $\times$ 1      & 128 $\times$ N $\times$ 32      & 544 \\
		MFM-2-1               & --                & 128 $\times$ N $\times$ 16      & -- \\
		BatchNorm-1           & --                & 128 $\times$ N $\times$ 16      & 512 \\
		Conv2D-2-2            & 3 $\times$ 3      & 128 $\times$ N $\times$ 64      & 10K \\
		MFM-2-2               & --                & 128 $\times$ N $\times$ 32      & -- \\
		MaxPooling-2          & 2 $\times$ 1      & 64  $\times$ N $\times$ 32      & --  \\
		\midrule
		Conv2D-3-1            & 1 $\times$ 1      & 64 $\times$ N $\times$ 64      & 2K  \\
		MFM-3-1               & --                & 64 $\times$ N $\times$ 32      & -- \\
		BatchNorm-2           & --                & 64 $\times$ N $\times$ 32      & 256 \\
		Conv2D-3-2            & 3 $\times$ 3      & 64 $\times$ N $\times$ 128     & 28K \\
		MFM-3-2               & --                & 64 $\times$ N $\times$ 64      & -- \\
		MaxPooling-3          & 2 $\times$ 1      & 32 $\times$ N  $\times$ 64     & --   \\
		\midrule
		Conv2D-4-1            & 1 $\times$ 1      & 32 $\times$ N $\times$ 96      & 5K  \\
		MFM-4-1               & --                & 32 $\times$ N $\times$ 48      & -- \\
		BatchNorm-3           & --                & 32 $\times$ N $\times$ 32      & 128 \\
		Conv2D-4-2            & 3 $\times$ 3      & 32 $\times$ N $\times$ 128     & 55K \\
		MFM-4-2               & --                & 32 $\times$ N $\times$ 64      & -- \\
		MaxPooling-4          & 2 $\times$ 1      & 16 $\times$ N $\times$ 64      & --   \\
		\midrule
		Conv2D-5-1            & 1 $\times$ 1      & 16 $\times$ N $\times$ 128     & 8K  \\
		MFM-5-1               & --                & 16 $\times$ N $\times$ 64      & -- \\
		BatchNorm-4           & --                & 16 $\times$ N $\times$ 64      & 64 \\
		Conv2D-5-2            & 3 $\times$ 3      & 16 $\times$ N $\times$ 160     & 92K \\
		MFM-5-2               & --                & 16 $\times$ N $\times$ 80      & -- \\
		MaxPooling-5          & 2 $\times$ 1      & 8  $\times$ N $\times$ 80      & --   \\
		\midrule
		Conv2D-6-1            & 1 $\times$ 1      & 8 $\times$ N $\times$ 192      & 13K  \\
		MFM-6-1               & --                & 8 $\times$ N $\times$ 96       & -- \\
		BatchNorm-5           & --                & 8 $\times$ N $\times$ 64       & 32 \\
		Conv2D-6-2            & 3 $\times$ 3      & 8 $\times$ N $\times$ 192      & 138K \\
		MFM-6-2               & --                & 8 $\times$ N $\times$ 96       & -- \\
		MaxPooling-6          & 2 $\times$ 1      & 4 $\times$ N $\times$ 96       & --   \\
		\midrule
		AttentionPooling      & --                & 4 $\times$ 96                  & 9K  \\
		Flatten               & --                & 384                            & --  \\
		\midrule
		Dense1                & --                & 256                            & 99K  \\
		Dense2                & --                & 256                            & 66K  \\
		Dense (softmax)       & --                & 10                             & 2570  \\
		\midrule
		Total                 & --                & --                             & 531K  \\
		\bottomrule
		\bottomrule
	\end{tabular}
	\vspace{-2mm}
\end{table}

\subsection{One-dimensional CNN for x-vector extraction}

In contrast to the other two proposed networks, the x-vector topology only uses one-dimensional convolution along the time. Table~\ref{tbl.xvector_topo} shows the network architecture. The network has three parts. The first part operates on the frame-by-frame level and outputs sequence of activation vectors (one for each frame). The second part compresses the frame-by-frame information into a fixed length vector of statistics describing the whole acoustic segment. More precisely, the weighted mean and weighted standard deviation of the input activation vectors are calculated over frames using the attention mechanism (note that in the original x-vector paper simple mean and standard deviation were used~\cite{snyder2018x}). The last part of the network consists of two Dense Leaky-ReLU layers followed by a Dense softmax layer like in the two previous topologies. Unlike our system for DCASE challenge 2018~\cite{zeinali2018convolutional} where we used the x-vector network in two ways: the softmax output was used for the classification or the x-vectors extracted at the output of the first affine transform after the pooling layer are used as the input for another classifier. Here we only use this network for classification exactly the same as the other two networks.

\begin{table}[!t]
  \renewcommand{\arraystretch}{1.1}
  \caption{\label{tbl.xvector_topo} 1-Dimensional CNN topology for x-vector extraction. The second column shows the relative indices to the current time step. N in the third column indicates the segment length which is 128 for the training phase and 512 for the evaluation phase. The attention layer here uses both mean and standard deviation statistics.}
  \vspace{2mm}
  \centering{
    \setlength\tabcolsep{6pt}
    \begin{tabular}{l c c r}
		\toprule
		\toprule
		\textbf{Layer name}   & \textbf{Filters Index}   & \textbf{Output}                   & \textbf{\#Params} \\
		\midrule
		Input                 & --                & N $\times$ 256       & -- \\
		Conv1D-1              & (-2,-1,0,1,2)     & N $\times$ 256       & 328K \\
		BatchNorm-1           & --                & N $\times$ 256       & 1K \\
		Dropout-1             & --                & N $\times$ 256       & -- \\
		\midrule
		Conv1D-2              & (-2,0,2)          & N $\times$ 256       & 197K \\
		BatchNorm-2           & --                & N $\times$ 256       & 1K \\
		Dropout-2             & --                & N $\times$ 256       & -- \\
		\midrule
		Conv1D-3              & (-3,0,3)          & N $\times$ 256       & 197K \\
		BatchNorm-3           & --                & N $\times$ 256       & 1K \\
		Dropout-3             & --                & N $\times$ 256       & -- \\
		\midrule
		Conv1D-4              & (-4,0,4)          & N $\times$ 256       & 197K \\
		BatchNorm-4           & --                & N $\times$ 256       & 1K \\
		Dropout-4             & --                & N $\times$ 256       & -- \\
		\midrule
		Conv1D-5              & (0)               & N $\times$ 256       & 66K \\
		BatchNorm-5           & --                & N $\times$ 256       & 1K \\
		Dropout-5             & --                & N $\times$ 256       & -- \\
		\midrule
		Conv1D-6              & (0)               & N $\times$ 768       & 197K \\
		BatchNorm-6           & --                & N $\times$ 768       & 1K \\
		Dropout-6             & --                & N $\times$ 768       & -- \\
		\midrule
		AttentionPooling      & --                & 1536                 & 590K \\
		\midrule
		Dense1                & --                & 256                  & 394K  \\
		BatchNorm-7           & --                & 256                  & 1K \\
		Dropout-7             & --                & 256                  & -- \\
		\midrule
		Dense2                & --                & 256                  & 66K  \\
		BatchNorm-8           & --                & 256                  & 1K \\
		Dense3 (softmax)      & --                & 10                   & 2560  \\
		\midrule
		Total                 & --                & --                   & 2240K  \\
		\bottomrule
		\bottomrule
	\end{tabular}
    \vspace{-5mm}
  }
\end{table}

\subsection{Attention Mechanism}

The conventional mean pooling layer considers the same weight for each input channel (depending on specified dimensions for calculating the mean). Each audio signal in the acoustic scene classification task contains several audio events which happened only in few frames in addition to the other events that happened in all frames like background noise. So, some frames contain more information than others about the interesting scene (i.e. class) and we should pay more attention to them. This is not possible with conventional mean pooling layer. We have proposed to use the attention mechanism which already successfully used in speaker verification task~\cite{zhu2018self,okabe2018attentive,chowdhury2017attention, zeinali2019improve}. In this method, the last layer before pooling layer is used to calculate weights of frames. Then the weighted mean and weighted standard deviation of the input channels are calculated and used as the output of the layer. For x-vector topology using both mean and standard deviation perform better in our setup while for the other two proposed topologies using only mean is slightly better. 


\section{Systems and Fusion}

In this challenge, we fused outputs of different networks to obtain the final results. First, we made a 4-folds cross-validation setup using the whole development data in addition to the official provided setup (we only use the official validation set for report results here and for the final system we only used the generated folds). By doing some initial experiments using the provided setup we found there are some easy and some difficult locations in the development data. In order to be able to make any valid conclusions, it is better to evaluate the networks on the whole development data. So, by following the proposed strategy in \cite{Dorfer2018}, we made this 4-folds validation setup.

For each fold, we trained 3 proposed network topologies using the data from the other three folds and evaluate them on the selected fold. The results on each fold were used to train a fusion system on the output of the trained networks (i.e. the output of the affine transform before applying softmax activation). FoCal Multiclass toolbox~\cite{brummer2007focal} was used for the training of a fusion system based on logistic regression. The final output of the system was the average fused output of each fold. Note that the output of FoCal is a calibrated score (i.e. have pretty same score distributions) and the average of them performs quite well.

As an alternative to fusion training, we also did a majority vote fusion. We have 12 trained networks for all folds. In this case, we first classify the test segment on all networks and count the number of classified output for all 10 classes. The final segment class was the most voted class. If two classes had the same number of votes, the class with the higher score from the first fusion method was used.

\section{Experiments and Results}

\subsection{Experimental Setups}

Similar to the baseline system provided by the organizers, our networks training was performed by optimizing the categorical cross-entropy using Adam optimizer~\cite{kingma2014adam}. The initial learning rate was set to 0.001 and the network training was early-stopped if the validation loss did not decrease for more than 100 epochs. The learning rate was linearly decreased to 1e-6 starting from epoch 50. The maximum number of epochs and the mini-batch size were set to 500 and 128, respectively.

\subsection{Results on the Official Fold}

In this section, the results of the final system with trained fusion for each scene are reported. Table~\ref{tbl.fold_results} shows the performance of the system for each scene separately as well as the overall performance on the official challenge validation fold.

\begin{table}[t]
  \renewcommand{\arraystretch}{1.1}
  \caption{\label{tbl.fold_results} Comparison results between different scenes of the final fused system.}
  \vspace{2mm}
  \centerline
  {
  \setlength\tabcolsep{8pt}
    \begin{tabular}{ l c c c c }
    \toprule
    \midrule
            			& & Our system      & Baseline \\
    Scene label			& & Accuracy [\%]   & Accuracy [\%]\\
    \midrule
    Airport				& & 71.5 &          48.4 \\
    Bus					& & 92.7 &          62.3 \\
    Metro				& & 74.3 &          65.1 \\
    Metro Station		& & 75.2 &          54.5 \\
    Park				& & 92.9 &          83.1 \\
    Public Square		& & 58.6 &          40.7 \\
    Shopping Mall		& & 71.8 &          59.4 \\
    Street Pedestrian	& & 60.0 &          60.9 \\
    Street Traffic		& & 90.6 &          86.7 \\
    Tram				& & 81.9 &          64.0 \\
    \midrule
    Average				& & 77.0 &          62.5 \\
    \midrule
    \bottomrule
    \end{tabular}
    \vspace{-4mm}
  }
\end{table}

\section{Conclusions}

We have described the systems submitted by BUT team to Acoustic Scene Classification (ASC) challenge of DCASE2019. Different systems were designed for this challenge and the final systems were fusions of the output scores from the individual system. A trained fusion as well as a majority vote fusion were used for the final system. The proposed systems are the fusion of there different network topologies: VGG like, Light-CNN and x-vector.



\bibliographystyle{IEEEtran}
\bibliography{refs}

\end{sloppy}
\end{document}